\documentclass[letterpaper]{emulateapj}

\usepackage{psfig}
\usepackage{rotating}
\usepackage{url}

\newcommand{\Chandra}{{\em Chandra}}
\newcommand{\chandra}{\textit{Chandra}}
\newcommand{\chandralong}{\textit{Chandra X-ray Observatory}}
\newcommand{\xmmlong}{\textit{XMM-Newton}}
\newcommand{\xmm}{\textit{XMM}}
\newcommand{\rosat}{{\em ROSAT}}

\newcommand{\nh}{\mbox{$N_{\rm H}$}} 
\newcommand{\nhtt}{\mbox{$N_{\rm H,22}$}}

\newcommand{\fxfj}{\mbox{$F_{\rm X}/F_{J}$}}

\newcommand{\kteff}{\mbox{$kT_{\rm eff}$}}
\newcommand{\rinfty}{\mbox{$R_{\infty}$}}
\newcommand{\rns}{\mbox{$R_{\rm NS}$}}
\newcommand{\mns}{\mbox{$M_{\rm NS}$}}

\newcommand{\chisq}{\mbox{$\chi^2$}}
\newcommand{\Chisq}[3]{$\chi^2_\nu$/dof (prob.) = {#1}/{#2} (#3)}
\newcommand{\chisqrnu}{$\chi^2_\nu$}

\newcommand{\Fx}{\mbox{$F_{\rm X}$}}

\newcommand{\xray}{\mbox{X-ray}}

\def\whereisnorth{In this figure, north is up and R.A. increases to the left}

\newcommand{\simlt}{\mathrel{\hbox{\rlap{\hbox{\lower4pt\hbox{$\sim$}}}\hbox{$<$}}}}
\newcommand{\simgt}{\mathrel{\hbox{\rlap{\hbox{\lower4pt\hbox{$\sim$}}}\hbox{$>$}}}}
\newcommand{\approxgt}{\mbox{$\,^{>}\hspace{-0.24cm}_{\sim}\,$}}

\newcommand{\ee}[1]{\mbox{$10^{#1}$}}
\newcommand{\tee}[1]{\mbox{$\times 10^{#1}$}}
\newcommand{\ud}[2]{\mbox{$^{+ #1}_{- #2}$}}
\newcommand{\ppm}{\mbox{$\pm$}}

\newcommand{\unit}[1]{\mbox{$\rm\,#1$}}
\def\deg{\hbox{$^\circ$}}
\def\arcdeg{\hbox{$^\circ$}}
\def\arcmin{\hbox{$^\prime$}}
\def\arcsec{\hbox{$^{\prime\prime}$}}
\def\sec{\mbox{$\,{\rm sec}$}}
\newcommand{\msun}{\mbox{$\,M_\odot$}}

\newcommand{\lsun}{\mbox{$\,L_\odot$}}

\newcommand{\km}{\hbox{$\,{\rm km}$}}

\newcommand{\MeV}{\mbox{$\,{\rm MeV}$}}
\newcommand{\keV}{\mbox{$\,{\rm keV}$}}
\newcommand{\eV}{\mbox{$\,{\rm eV}$}}
\newcommand{\ksec}{\mbox{$\,{\rm ks}$}}

\newcommand{\kpc}{\mbox{$\,{\rm kpc}$}}
\newcommand{\pc}{\mbox{$\,{\rm pc}$}}
\newcommand{\persec}{\mbox{$\,{\rm s^{-1}}$}}
\newcommand{\perksec}{\mbox{$\,{\rm ks^{-1}}$}}

\newcommand{\percmsq}{\mbox{$\,{\rm cm^{-2}}$}}
\newcommand{\percmcube}{\mbox{$\,{\rm cm^{-3}}$}}
\newcommand{\peryear}{\mbox{$\,{\rm yr^{-1}}$}}

\newcommand{\cgsflux}{\mbox{$\,{\rm erg\,\percmsq\,\persec}$}}
\newcommand{\cgslum}{\mbox{$\,{\rm erg\,\persec}$}}
\newcommand{\cgsdensity}{\mbox{$\,{\rm g\percmcube}$}} 
\newcommand{\cgsaccel}{\mbox{$\,{\rm cm\,s^{-2}}$}} 
\newcommand{\masperyr}{\mbox{$\,{\rm mas\peryear}$}}

\def\sourcefour{XMMU~J171433$-$292747}
\def\sourcethree{XMMU~J180916$-$255425}
\def\sourcenineB{XMMU~J180839$-$260118}
\def\TwoMassnineB{2MASS~J18083961$-$2601203} 

\begin{document}

\title{Discovery of a candidate quiescent low-mass X-ray binary in
the globular cluster NGC~6553}

\author{Sebastien Guillot\footnote{Vanier Canada Graduate Scholar}, Robert E. Rutledge}
\affil{Department of Physics, McGill University,\\ 3600 rue
  University, Montreal, QC, Canada, H3A-2T8}
\email{guillots@physics.mcgill.ca}

\author{Edward F. Brown} \affil{Department of Physics and Astronomy,
  Michigan State University, 3250 Biomedical Physical Science
  Building, East Lansing, MI 48824-2320, USA} 

\author{George G. Pavlov} \affil{Pennsylvania State University, 512
  Davey Lab, University Park, PA 16802, USA} \affil{St.-Petersburg
  State Polytechnical University, Polytechnicheskaya ul. 29,
  St.-Petersburg 195251, Russia}

\author{Vyacheslav E. Zavlin} \affil{Space Science Laboratory,
  Universities Space Research Association, NASA MSFC VP62, Huntsville,
  AL 35805, USA}

\slugcomment{Draft submitted to ApJL on \today}
\shorttitle{Discovery of qLMXBs in NGC~6553}

\begin{abstract}
This paper reports the search for quiescent low-mass \xray\ binaries
(qLMXBs) in the globular cluster (GC) NGC~6553 using an
\xmmlong\ observation designed specifically for that purpose.  We
spectrally identify one candidate qLMXB in the core of the cluster,
based on the consistency of the spectrum with a neutron star
H-atmosphere model at the distance of NGC~6553.  Specifically, the
best-fit radius found using the three \xmm\ European Photon Imaging
Camera spectra is $\rns=6.3\ud{2.3}{0.8}\km$ (for $\mns=1.4\msun$) and
the best-fit temperature is $\kteff=136\ud{21}{34}\eV$.  Both physical
parameters are in accordance with typical values of previously
identified qLMXBs in GC and in the field, i.e., $\rns\sim5$--$20\km$
and $\kteff\sim50$--$150 \eV$.  A power-law (PL) component with a
photon index $\Gamma=2.1\ud{0.5}{0.8}$ is also required for the
spectral fit and contributes $\sim 33\%$ of the total flux of the
\xray\ source.  A detailed analysis supports the hypothesis that the
PL component originates from nearby sources in the core, unresolved
with \xmm.  The analysis of an archived \chandra\ observation provides
marginal additional support to the stated hypothesis.  Finally, a
catalog of all the sources detected within the \xmm\ field of view is
presented here.

\end{abstract}

\maketitle



\section{Introduction}

The faint \xray\ emission ($L_{\rm X}\sim10^{32}$--$10^{33}\cgslum$)
of transiently accreting low-mass \xray\ binaries in quiescent state
(qLMXBs) was historically interpreted as a thermal blackbody with
emission area smaller than that expected for a 10 km neutron star
(NS).  It was later claimed that the observed luminosity was not due
to low accretion rate onto the NS surface, as initially suggested
\citep{vanparadijs87}, but due to heat from the NS crust radiating
through the upper layers of the NS atmosphere \citep{brown98}.  In
this interpretation, called Deep Crustal Heating, the energy is
deposited in the NS crust by pressure-sensitive nuclear reactions
(electron captures, neutron emissions and pycnonuclear reactions,
\citealt{gupta07, haensel08}), as matter accumulates on the surface
during episodes of rapid accretion.  This chain of reactions, from the
NS surface to depths with the density of undifferentiated equilibrium,
releases $\sim 1.5\MeV$ per accreted nucleon in the crust.  In a
steady state, these reactions give rise to a time-averaged luminosity,
which is directly proportional to the time-averaged mass accretion rate
\citep{brown98}:
\begin{equation}
 \langle L \rangle= 9\tee{32}\,\frac{\langle \dot{M}
    \rangle}{10^{-11}\unit{\msun\,
    \peryear}}\,\frac{Q}{1.5\unit{MeV/amu}}\cgslum
  \label{eq:dch}
\end{equation}
\noindent where $Q$ is the average heat deposited in the NS crust per
accreted nucleon.

The resulting thermal spectrum is described by a realistic NS
atmosphere composed exclusively of hydrogen.  Indeed, the
gravitational settling of the accreted material from the low-mass
companion star happens on time scales of $\sim$seconds
\citep{bildsten92}, resulting in a pure H-atmosphere around the NS.
The current models of NS H-atmosphere \citep{zavlin96, mcclintock04,
  heinke06a} show that the observed emission is consistent with that
from the entire surface area of a $\sim$10 km NS \citep{rutledge99}.
The H-atmosphere interpretation allows observers to determine the
physical radius $R_{\rm NS}$ (or the projected radius $\rinfty =
R_{\rm NS} \left(1+z\right)$, with $1+z=\left(1-2G\mns/\rns
c^{2}\right)^{-1/2}$) using spectral fitting in the soft \xray\ range,
where the thermal spectrum of qLMXBs peaks.

The radius measurement of NSs using the thermal spectrum of qLMXBs is
one way to place constraints on the equation of state (EoS) of dense
nuclear matter.  This relation between the pressure and the density in
the interior of NSs is highly uncertain  and multiple competing
theories are proposed to explain the behavior of cold matter at
densities above $2.35\tee{14}\cgsdensity$.  Quiescent LMXBs found in
globular clusters (GCs) are routinely used to produce precise
measurements of NS radii.  The distance to GCs are generally known
with precisions of the order of $\sim 5\%$--$10\%$  and the
overabundances of NS binary systems in GCs \citep{hut92} make these
objects ideal targets to pursue searches of qLMXBs capable of
providing useful constraints on the dense matter EoS.

Twenty-six GC qLMXBs, including the more recent candidates, have been
discovered so far \citep[see][for complementary
  lists]{heinke03a,guillot09a}.  However, for some of them, the poorly
constrained \rns\ and \kteff\ measurements, as well as the high
galactic absorption in their direction (for example,
$\nh\sim1.2\tee{22}\unit{atoms\percmsq}$ in the direction of
Terzan~5), affects the certainty of their identification.  While
qLMXBs in the field of the Galaxy or in GC have been observed in
quiescence following outbursts events in historical \xray\ transients,
none of the spectrally identified qLMXBs in GC (with spectra
consistent with NS atmosphere models or via their X-ray spectral
colors) have been observed in outbursts.  There has been one exception
recently discovered.  An \xray\ transient in the cluster Terzan~5 has
been observed in outburst \citep{bordas10,pooley10,degenaar11} after
being tentatively classified as a qLMXB \citep[CX25,][]{heinke06b}.
To increase the list of known GC qLMXBs, a program of short
observations using the European Photon Imaging Cameras
\citep[EPIC;][]{struder01} onboard \xmmlong\ have been undertaken to
survey GCs and search for qLMXBs.  This paper reports the discovery
via spectral identification of a candidate qLMXB in the core of
NGC~6553.  A short archived \chandra\ observation targeted at this GC
was also analyzed to provide more accurate positions and tentatively
confirm the source classification.

The targeted cluster, NGC~6553, is a low-galactic latitude GC located
at the position R.A.=18$^{\rm h}$09$^{\rm m}$17.6$^{\rm s}$ and ${\rm
  decl.}=-25\deg 54\arcmin 31\farcs3$ (J2000), approximately
2.2\kpc\ from the galactic center, corresponding to a heliocentric
distance of $d=6.0\kpc$.  It is a GC of moderate core compactness and
moderate size: core radius $r_{\rm c}=0\farcm55$, half mass radius
$r_{\rm HM}= 1\farcm55$ and tidal radius $r_{\rm t}=8\farcm16$.  Its
metallicity $\left[{\rm M/H}\right] = -0.09$ \citep{valenti07} makes
it one of the most metallic clusters in the Galaxy.  The foreground
reddening $E\left(B-V\right)=0.63$ corresponds to a moderately high
hydrogen column density $\nh=0.35\tee{22} \unit{atoms\percmsq}$
\citep{predehl95}.  The hydrogen column density is written
$\nhtt=0.35$ afterward, and this value will be used for the
\xray\ spectral fits.  This value is similar to the value found from
NRAO data \citep{dickey90}\,\footnote {From
  \url{http://cxc.harvard.edu/toolkit/colden.jsp}}, which accounts for
the atomic hydrogen $N_{\rm HI}$ through the entire galaxy.  The GC
properties come from the catalog of GCs \citep[][update
  Dec. 2010]{harris96}, except when other references are provided.

The organization of this paper is as follows. Section~\ref{sec:red}
describes the data reduction and analysis of the \xmmlong\ and
\chandra\ observations.  In Section~\ref{sec:res6553}, we present the
results and we discuss them and conclude in Section~\ref{sec:discuss}

\section{Data Reduction and Analysis}
\label{sec:red}

NGC~6553 was observed with \xmmlong\ on 2006 October 6 at 01:41:44
UT, using the three EPIC cameras, for an exposure time of 20.4\ksec,
with the medium filter.  The \chandralong\ observed the GC for
5.25\ksec, on 2008 October 30 at 01:38:12 UT, with the ACIS-S
detector \citep{garmire03} in VERY-FAINT mode.

\subsection{Data reduction}
\label{sec:reduction}

Reduction of \xmmlong\ data is performed with the \xmmlong\ Science
Analysis Software (SAS) v8.0.0, using the standard
procedures\footnote{User Guide to the \xmmlong\ Science Analysis
  System, Issue 5.0, 2008 (ESA: \xmmlong\ SOC).}.  More
specifically, {\tt epchain} and {\tt emchain} are used for the
preliminary reduction of the raw pn and MOS data files, respectively.
Single and double patterns are used for pn data and single, double and
quadruple patterns for the MOS cameras, both in the
0.3--10\keV\ range.  Prior to the source detection, the data are
checked for background flares by looking for $>3\sigma$ deviation from
the mean count rate of the entire detector.  None are found and the
whole integration time is used.  Also, an exposure map is created to
correct for the effect of vignetting.  The script {\tt wavdetect} from
CIAO v4.2\ \citep{fruscione06} is then run on the pn image in the
0.3--10\keV\ range, with the following parameters: minimum relative
exposure of 0.1, the wavelet scales {\it ``1.0 2.0 4.0 8.0''} and a
significance threshold $3\tee{-6}$.  The latter is equal to the
inverse of the number of pixels in the image, and allows for about one
spurious detection in the entire image.  Sources with detection
significance $\sigma > 4$ are retained for analysis.

The reduction and analysis of the \chandra\ data are accomplished with
the CIAO v4.2 package \citep{fruscione06}.  The level-1 event file is
first reprocessed using the public script \emph{chandra\_repro} which
performs the steps recommended by the data preparation analysis
thread\footnote{\url{http://cxc.harvard.edu/ciao/threads/data.html}}
(corrections for charge transfer inefficiency, destreaking, bad pixel
removal, etc, if needed) making use of the latest effective area maps,
quantum efficiency maps, and gain maps of CALDB v4.3
\citep{graessle07}.  The newly created level-2 event file is then
checked for background flares, but there are none detected.

\subsection{Count extraction}
\label{sec:spectra}

The command {\tt evselect} from the SAS is used for source and
background count extraction of the \xmm\ data.  A radius of
30\arcsec\ around each source is chosen, accounting for 88\% of the
total energy from an on-axis source at 1.5\keV\ with the pn
detector\footnote{From \xmmlong\ Users Handbook, Figure 3.7, July
  2010}.  The background counts, used for background subtraction, are
extracted from a larger region (100\arcsec) around the source of
interest, excluding the source itself and other detected sources in
close proximity with a circular region of 40\arcsec\ around each of
them, ensuring that 92\% of the overlapping source counts (at
1.5\keV\ on the pn detector) are excluded.  In some cases, the close
proximity of the detected sources requires to adapt the exclusion
radii of nearby sources.  Those few cases are discussed in
Section~\ref{sec:res6553}.  The tasks {\tt rmfgen} and {\tt arfgen}
generate the response matrix files (RMF) and the ancillary response
files (ARF) for each observations.  The \xmmlong\ extracted counts are
binned using {\tt grppha} into energy bins between 0.3 and
10\keV\ with a minimum of 20 counts per bin.  This criterion ensures
having approximate Gaussian statistics in each spectral bin.

For \Chandra\ data, the script {\tt psextract} is used to extract the
counts of the \xray\ sources, in the energy range 0.5--8\keV.  The
extraction region depends on the off-axis angle since the point spread
function (PSF) degrades at large off-axis angles.  Since the analyzed
observation of NGC~6553 was performed with the ACIS-S instrument and a
focal plane temperature of -120\unit{\deg C}, the RMFs have to be
recalculated, according to the recommendations of the CIAO Science
Thread \emph{``Creating ACIS RMFs with mkacisrmf''}. It is also
crucial to recalculate the ARFs using the new RMFs in order to match
the energy grids between the two files.

In both the \xmmlong\ and \chandra\ observations, the count rates are
low enough for the effects of pile-up to be ignored.  Specifically,
for the \chandra\ data, the count rate was 0.031 photons per detector
frame (3.24\sec), and therefore the effect of pile-up can be safely
neglected.

\subsection{Spectral Analysis}
\label{sec:analysis}

The identification of candidate qLMXBs in this analysis is based on
the spectral identification with the models that empirically describe
the spectra of previously known qLMXBs.  Spectra from the pn camera
are first extracted and analyzed, since its superior efficiency
facilitates spectral analysis with an improved signal-to-noise ratio
(S/N) over the two MOS cameras.  H-atmosphere models are available in
\emph{XSPEC} \citep{arnaud96}, including {\tt nsa} and {\tt nsagrav}
\citep{zavlin96} and {\tt nsatmos} \citep{mcclintock04, heinke06a}.
The major difference between the models resides in the surface gravity
values $g$ used to calculate the model: {\tt nsa} has been implemented
with a fixed value $g=2.43\tee{14}\cgsaccel$, {\tt nsagrav} has been
computed for a range $g=(0.1$--$10)\tee{14} \cgsaccel$ and {\tt
  nsatmos} for the range $g=(0.63$--$6.3)\tee{14} \cgsaccel$.
Previous works have demonstrated that NS atmosphere models calculated
with multiple values of $g$ are better adapted to fit the thermal
spectra of qLMXBs \citep{heinke06a,webb07}.  We choose here to use
{\tt nsatmos}.

The identification of candidate qLMXBs is based on the statistical
consistency (null hypothesis probability ${\rm n.h.p}\approxgt
\ee{-2}$) with the {\tt nsatmos} model at the distance of the host GC,
with the galactic absorption taken into account using the
multiplicative model {\tt wabs}.  Additional conditions impose that
the best-fit effective temperature and projected radius are within the
range of previously observed GC qLMXBs, i.e., $\kteff\sim50$--$180\eV$
and $\rinfty \sim 5$--$20 \km$ \citep[see][for complementary lists of
  GC qLMXBs]{heinke03c,guillot09a}.  For identified candidate qLMXBs,
the pn, MOS1 and MOS2 extracted spectra are then fit simultaneously to
improve statistics and diminish the uncertainties on the best-fit
parameters.

In some cases, an excess of counts at high energy is modeled with an
additional power-law (PL) component.  This is empirically justified by
the fact that qLMXBs sometimes display a hard PL which dominates the
spectrum above 2\keV.  Most qLMXBs in the field of the galaxy have
spectra best fitted with H-atmosphere models combined with a
hard-photon PL (for example, Cen~X-4, \citealt{asai96,rutledge01a},
Aql~X-1, \citealt{asai98,campana98,rutledge01b}).  It has been claimed
that some GC qLMXBs also displayed significant PL contributions
(between $\sim10\%$ and $\sim50\%$ of the total flux), for example, in
NGC~6440 \citep{cackett05}, and in Ter~5 \citep{heinke06b}.  Proposed
interpretations of the observed PL tail include residual accretion
onto the NS magnetosphere \citep{grindlay01a,cackett05}, shock
emission via the emergence of a magnetic field \citep{campana00b}, or
an intrabinary shock between the winds from the NS and its companion
star \citep{campana04b}.  However, analyses of the quiescent emission
of LMXBs have shown evidence of the presence of a variable low-level
accretion on the NS; for the LMXB Aql~X-1 \citep{rutledge02a}, for
XTE~J1701$-$462, \citep{fridriksson10}, and for the LMXB Cen~X-4
\citep{cackett10}.

Such an additional spectral component for qLMXBs in the core of GCs
might not necessarily originate from one of the interpretations listed
above.  For example, a PL component was required in addition to the
H-atmosphere model for the spectral fit of the candidate qLMXB
\sourcefour\ in NGC~6304 observed with \xmm\ \citep{guillot09a}.
However, a short $5\ksec$ archived \chandra\ observation demonstrated
that the high energy excess of counts noticed for this source
originated from another close by \xray\ source (possibly a cataclysmic
variable) unresolved in the \xmm\ data \citep{guillot09b}.  Then, a
longer $100\ksec$ \chandra\ exposure confirmed the NS radius
measurement and the initially reported results (S. Guillot et
al. 2011, in preparation).

The main focus of this paper is the discovery of candidate qLMXBs in
NGC~6553, but as part of the systematic analysis of \xmm\ data, a
simple spectral analysis is performed for all other \xray\ sources
that are not consistent with NS H-atmosphere models at the distance of
the host GC.  Specifically, a simple absorbed PL model is used and
their best-fit photon indices, as well as their unabsorbed flux
(0.5--10\keV) are reported.  In some cases, the \xray\ spectra cannot
be fit adequately with either an NS H-atmosphere model or a PL model.
In those cases, the use of other models is attempted for completeness
and the details are provided in Section~\ref{sec:res6553}.  Finally,
sources with an $S/N<3$ are fit with a fixed ($\Gamma = 1.5$) photon
index.

In all cases, the galactic absorption is taken into account with the
model {\tt wabs}, keeping the value of \nh\ fixed at the value in the
direction of the GC, i.e., $\nhtt=0.35$.  Also, all quoted
uncertainties on the spectral parameters are 90\% confidence level in
the text and the tables.  Finally, the quoted fluxes are corrected for
the finite aperture (extraction radius), except noted otherwise.

\subsection{Astrometric correction}
\label{sec:astro}

The uncertainties reported in Table~\ref{tab:sources6553} on the
\xmm\ positions are composed of the {\tt wavdetect} statistical error
on the pn positions (up to 1\arcsec) and the systematic uncertainty of
\xmm\ (2\arcsec at $1\sigma$\,\footnote{from \xmm\ Science Operations
  Centre XMM-SOC-CAL-TN-0018, \cite{guainizzi08}}).  An absolute
astrometric correction -- into the frame of the Two Micron All Sky
Survey (2MASS) -- is applied through the association of \xray\ sources
with 2MASS counterparts.  Three \xray\ sources were firmly associated
with 2MASS objects.  The sources \#5, \#2 and \#9 have probabilities
0.001\%, 0.21\% and 0.04\%, respectively, that a star as bright or
brighter as the stars 2MASS~J18093564$-$25555377,
2MASS~J18100723$-$25511760 and \TwoMassnineB, respectively, lie as
close or closer to the three \xray\ sources.  These probabilities were
calculated using all 2MASS stars detected in 2\arcmin\ wide annulli,
centered at the GC optical center, and with respective average radii
equal to the distances of the respective 2MASS counterparts from the
GC center.  This means that sources \#5, \#2 and \#9 have
probabilities 99.999\%, 99.79\% and 99.96\% to be associated with the
stars 2MASS~J18093564$-$25555377, 2MASS~J18100723$-$25511760 and
\TwoMassnineB, respectively.  From these three associations, the
best-fit required transformation is a shift of -0\farcs1 in R.A. and
0\farcs71 in decl.  The residual uncertainties are 0\farcs34 in
R.A. and 1\farcs08 in decl.  This is added in quadrature to the 2MASS
astrometric uncertainty ($\sim0\farcs15$ \footnote{Section 2 of
  \url{http://www.ipac.caltech.edu/2mass/releases/allsky/doc/}}) and
to the statistical uncertainty of the \xray\ source detection.

\begin{figure}[t]
    \centerline{~\psfig{file=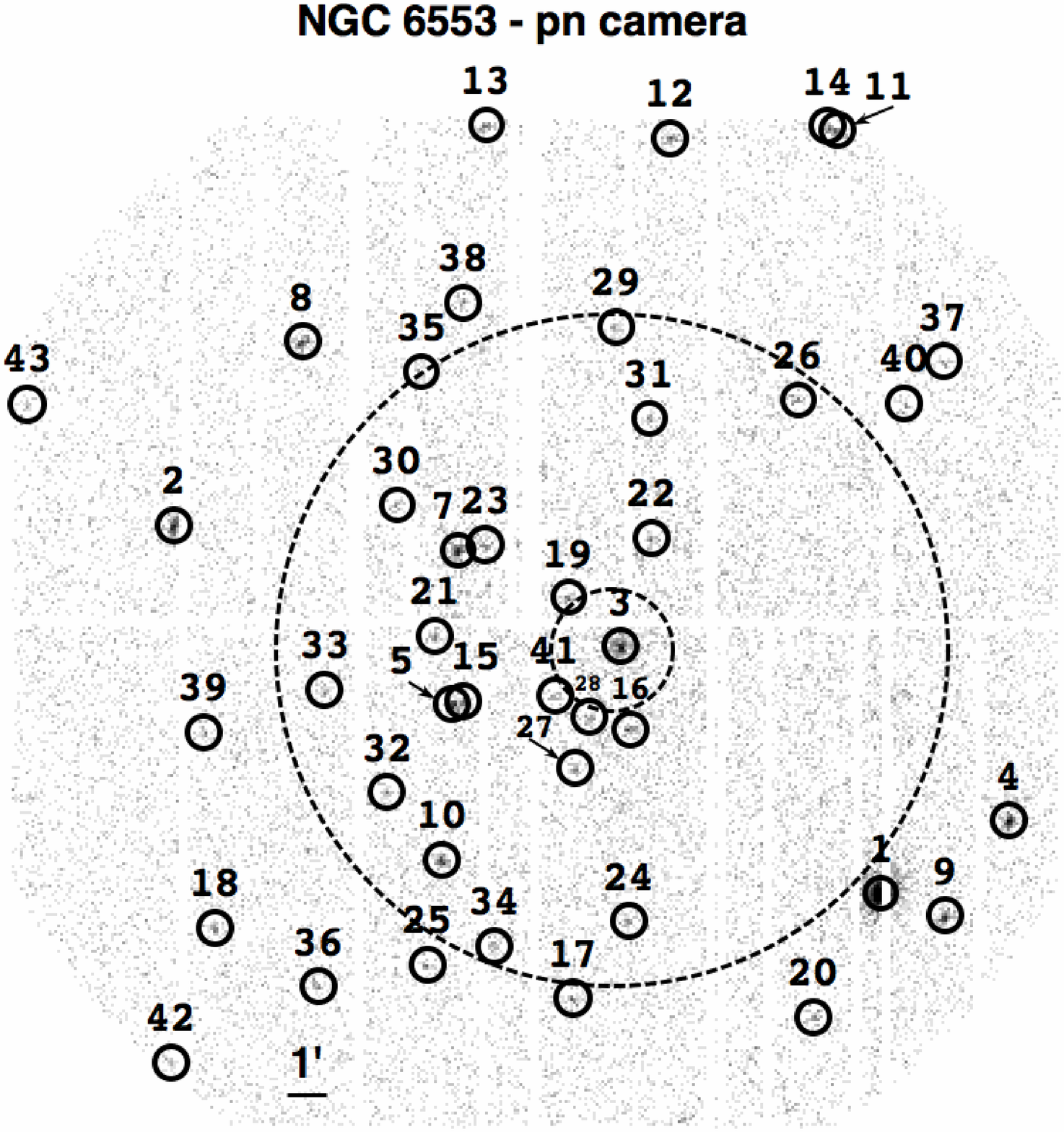,width=8.5cm,angle=0}~}
    \bigskip
    \caption[]{\label{fig:6553} This EPIC/pn image of NGC~6553 shows
      the 43 sources detected on the pn camera.  The small and large
      dashed circles respectively represent the half-mass ($r_{\rm
        HM}=1\farcm55$) and tidal radii ($r_{\rm t}=8\farcm16$).  A
      log gray-scale has been used for the image due to the brightness
      of source \#1.  \whereisnorth.  The candidate qLMXB detected in
      this GC is the source \#3, corresponding to \sourcethree.}
\end{figure}

\section{Results}
\label{sec:res6553}

\begin{deluxetable*}{rrrcrr}[t]
  \tablecaption{\label{tab:sources6553} \xray\ sources detected in the \xmm\ observation of NGC~6553.}
  \tablewidth{0pt}
  \tabletypesize{\scriptsize}    
  \tablecolumns{6}
  \tablehead{
    \colhead{Object Name} & \colhead{RA} & \colhead{decl.} &
    \colhead{$\delta_{\rm R.A.}\backslash \delta_{\rm decl.}$\tablenotemark{a} } & 
    \colhead{S/N\tablenotemark{b}} & \colhead{ID} \\
    \colhead{} & \colhead{(J2000)} & \colhead{(J2000)} & \colhead{(\arcsec)} & 
    \colhead{} & \colhead{} }
  \startdata
  XMMU~J180846$-$260043 & 272.19551 & -26.01220 & $\pm$0.0$\backslash$0.0 & 166.1 & 01\\ 
  XMMU~J181007$-$255118 & 272.53028 & -25.85518 & $\pm$0.3$\backslash$0.3 & 28.0 & 02\\ 
  XMMU~J180916$-$255425 & 272.31889 & -25.90704 & $\pm$0.4$\backslash$0.3 & 26.2 & 03\\ 
  XMMU~J180832$-$255852 & 272.13472 & -25.98135 & $\pm$0.4$\backslash$0.4 & 24.6 & 04\\ 
  XMMU~J180935$-$255554 & 272.39844 & -25.93172 & $\pm$0.4$\backslash$0.5 & 22.6 & 05\\ 
  XMMU~J180845$-$260035 & 272.18903 & -26.00978 & $\pm$0.4$\backslash$0.4 & 21.4 & 06\\ 
  XMMU~J180935$-$255157 & 272.39592 & -25.86605 & $\pm$0.3$\backslash$0.3 & 21.3 & 07\\ 
  XMMU~J180952$-$254637 & 272.46964 & -25.77697 & $\pm$0.7$\backslash$0.6 & 16.9 & 08\\ 
  XMMU~J180839$-$260118 & 272.16497 & -26.02191 & $\pm$0.6$\backslash$0.5 & 16.7 & 09\\ 
  XMMU~J180936$-$255952 & 272.40357 & -25.99803 & $\pm$0.6$\backslash$0.5 & 15.5 & 10\\ 
  XMMU~J180851$-$254113 & 272.21633 & -25.68706 & $\pm$0.6$\backslash$0.5 & 14.1 & 11\\ 
  XMMU~J180910$-$254125 & 272.29530 & -25.69036 & $\pm$0.5$\backslash$0.5 & 12.1 & 12\\ 
  XMMU~J180931$-$254105 & 272.38234 & -25.68474 & $\pm$0.6$\backslash$0.4 & 10.9 & 13\\ 
  XMMU~J180853$-$254106 & 272.22092 & -25.68513 & $\pm$0.6$\backslash$0.5 & 9.3 & 14\\ 
  XMMU~J180934$-$255549 & 272.39324 & -25.93030 & $\pm$0.5$\backslash$0.4 & 9.2 & 15\\ 
  XMMU~J180915$-$255631 & 272.31424 & -25.94212 & $\pm$0.7$\backslash$0.5 & 9.0 & 16\\ 
  XMMU~J180921$-$260325 & 272.34126 & -26.05712 & $\pm$1.0$\backslash$0.7 & 8.0 & 17\\ 
  XMMU~J181002$-$260135 & 272.51131 & -26.02663 & $\pm$0.6$\backslash$0.7 & 7.9 & 18\\ 
  XMMU~J180922$-$255310 & 272.34343 & -25.88638 & $\pm$0.9$\backslash$0.6 & 7.5 & 19\\ 
  XMMU~J180854$-$260353 & 272.22729 & -26.06497 & $\pm$0.6$\backslash$0.5 & 7.4 & 20\\ 
  XMMU~J180937$-$255408 & 272.40664 & -25.90245 & $\pm$0.5$\backslash$0.5 & 7.3 & 21\\ 
  XMMU~J180912$-$255138 & 272.30407 & -25.86082 & $\pm$0.8$\backslash$0.8 & 6.6 & 22\\ 
  XMMU~J180931$-$255149 & 272.38286 & -25.86367 & $\pm$0.8$\backslash$0.6 & 6.4 & 23\\ 
  XMMU~J180915$-$260127 & 272.31512 & -26.02427 & $\pm$0.9$\backslash$0.9 & 6.3 & 24\\ 
  XMMU~J180938$-$260233 & 272.41026 & -26.04259 & $\pm$1.0$\backslash$0.6 & 6.2 & 25\\ 
  XMMU~J180856$-$254806 & 272.23482 & -25.80185 & $\pm$0.9$\backslash$0.7 & 6.2 & 26\\ 
  XMMU~J180921$-$255731 & 272.34030 & -25.95878 & $\pm$0.8$\backslash$0.7 & 6.2 & 27\\ 
  XMMU~J180920$-$255614 & 272.33388 & -25.93725 & $\pm$0.7$\backslash$0.6 & 5.9 & 28\\ 
  XMMU~J180917$-$254614 & 272.32096 & -25.77083 & $\pm$0.9$\backslash$0.7 & 5.8 & 29\\ 
  XMMU~J180941$-$255048 & 272.42476 & -25.84671 & $\pm$0.9$\backslash$0.8 & 5.7 & 30\\ 
  XMMU~J180913$-$254834 & 272.30522 & -25.80972 & $\pm$0.9$\backslash$0.8 & 5.6 & 31\\ 
  XMMU~J180943$-$255808 & 272.42940 & -25.96905 & $\pm$0.7$\backslash$0.7 & 5.6 & 32\\ 
  XMMU~J180950$-$255531 & 272.45886 & -25.92548 & $\pm$0.7$\backslash$0.6 & 5.5 & 33\\ 
  XMMU~J180930$-$260205 & 272.37870 & -26.03489 & $\pm$0.8$\backslash$0.8 & 5.5 & 34\\ 
  XMMU~J180939$-$254723 & 272.41319 & -25.78992 & $\pm$0.0$\backslash$0.0 & 5.4 & 35\\ 
  XMMU~J180950$-$260307 & 272.46218 & -26.05219 & $\pm$0.9$\backslash$0.7 & 5.4 & 36\\ 
  XMMU~J180839$-$254707 & 272.16615 & -25.78536 & $\pm$0.7$\backslash$0.6 & 5.3 & 37\\ 
  XMMU~J180934$-$254537 & 272.39339 & -25.76053 & $\pm$0.6$\backslash$0.6 & 5.2 & 38\\ 
  XMMU~J181003$-$255636 & 272.51627 & -25.94347 & $\pm$0.7$\backslash$0.6 & 5.0 & 39\\ 
  XMMU~J180844$-$254812 & 272.18444 & -25.80336 & $\pm$0.8$\backslash$0.7 & 4.6 & 40\\ 
  XMMU~J180924$-$255540 & 272.35017 & -25.92790 & $\pm$0.7$\backslash$0.6 & 4.5 & 41\\ 
  XMMU~J181007$-$260503 & 272.53191 & -26.08422 & $\pm$0.7$\backslash$1.0 & 4.4 & 42\\ 
  XMMU~J181023$-$254811 & 272.59968 & -25.80325 & $\pm$0.7$\backslash$0.7 & 4.2 & 43\\ 
  \enddata 
  \tablenotetext{a}{ Statistical uncertainty on the position}
  \tablenotetext{b}{ Detection significance provided by {\tt wavdetect}}
\end{deluxetable*}

\begin{deluxetable*}{rrccrrr}[t]
  \tablecaption{\label{tab:spec6553} Spectral results of \xray\ sources in NGC~6553.}
  \tablewidth{0pt}
  \tabletypesize{\scriptsize}    
  \tablecolumns{7}
  \tablehead{
    \colhead{ID} & \colhead{pn Rate}  & \colhead{$\Gamma$} & \colhead{\kteff} &
    \colhead{$F_{\rm X}$} & \colhead{\chisqrnu/d.o.f. (prob.)} & \colhead{Model} \\
    \colhead{} & \colhead{(cts ks$^{-1}$)}  & \colhead{} & \colhead{(keV)} &
    \colhead{} & \colhead{} & \colhead{Used}}
  \startdata
  01 & 197.3\ppm3.8 & -- & 0.35\ud{0.05}{0.12} ; 2.80\ud{5.75}{0.67} & 17.3  & 1.01/115 (0.44) & RS+RS\\ 
  02 & 17.7\ppm1.3 & 3.0\ud{0.3}{0.3} & -- & 2.54 & 1.32/15 (0.18) & PL\\ 
  03 & 27.1\ppm1.5 \tablenotemark{a} & 2.1\ud{0.5}{0.8} & 0.136\ud{0.021}{0.034} & 1.29 & 0.90/54 (0.68) & NSATMOS+PL\\ 
  04 & 18.1\ppm1.4 & 1.4\ud{0.2}{0.2} & -- & 2.67 & 0.79/18 (0.71) & PL\\ 
  05 & 15.0\ppm1.3 & -- &  0.37\ud{0.07}{0.12} & 0.42 & 0.75/13 (0.71) & RS\\ 
  06 & 8.6\ppm0.9 & 3.3\ud{0.5}{0.4} & -- & 5.89 & 1.50/ 7 (0.16) & PL\\ 
  07 & 17.6\ppm1.4 & 0.9\ud{0.2}{0.2} & -- & 2.46 & 0.75/18 (0.76) & PL\\ 
  08 & 10.8\ppm1.1 & 0.8\ud{0.3}{0.3} & -- & 2.72 & 0.94/11 (0.50) & PL\\ 
  09 & 11.6\ppm1.2 & -- & 0.873\ud{0.145}{0.092} & 0.65 & 1.20/30 (0.21) & RS\\ 
  10 & 9.6\ppm1.1 & 2.4\ud{0.5}{0.4} & -- & 0.74 & 0.75/12 (0.70) & PL\\ 
  11 & 6.0\ppm0.9 & 0.7\ud{0.3}{0.4} & -- & 2.58 & 0.85/ 6(0.53) & PL\\ 
  12 & 4.0\ppm1.0 & -0.3\ud{0.7}{0.7} & -- & 2.44 & 0.83/ 8 (0.58) & PL\\ 
  13 & 3.9\ppm0.9 & 1.2\ud{0.6}{0.6} & -- & 1.07 & 0.92/ 6 (0.48) & PL\\ 
  14 & 3.7\ppm0.7 & 0.2\ud{0.6}{0.7} & -- & 2.2 & 0.66/ 3 (0.57) & PL\\ 
  15 & 6.7\ppm1.0 & 2.8\ud{0.8}{0.8} & -- & 0.44 & 2.02/ 8 (0.04) & PL\\ 
  16 & 4.5\ppm1.0 & 4.5\ud{1.5}{1.3} & -- & 0.31 & 1.27/ 9 (0.25) & PL\\ 
  17 & 4.3\ppm0.9 & 3.9\ud{1.2}{1.1} & -- & 0.34 & 2.15/ 7 (0.03) & PL\\ 
  18 & 5.1\ppm1.0 & 1.5\ud{0.5}{0.5} & -- & 0.96 & 0.59/ 8 (0.79) & PL\\ 
  19 & 3.7\ppm1.0 & 1.3\ud{0.7}{0.7} & -- & 0.30 & 0.86/ 9 (0.56) & PL\\ 
  20 & 4.0\ppm0.9 & 5.0\ud{1.4}{1.2} & -- & 0.51 & 1.30/ 7 (0.25) & PL\\ 
  21 & 4.4\ppm0.9 & 3.5\ud{1.0}{0.9} & -- & 0.31 & 1.14/ 7 (0.33) & PL\\ 
  22 & 2.3\ppm0.9 & (1.5) & -- & 0.13 & 0.54/ 8 (0.82) & PL\\ 
  23 & 4.6\ppm1.0 & 2.1\ud{0.7}{0.6} & -- & 1.60 & 0.65/ 8 (0.74) & PL\\ 
  24 & 3.5\ppm0.9 & 1.0\ud{0.6}{0.6} & -- & 0.47 & 1.31/ 7 (0.24) & PL\\ 
  25 & 4.4\ppm0.9 & 1.4\ud{1.7}{1.1} & -- & 0.44 & 1.84/ 7 (0.08) & PL\\ 
  26 & 2.6\ppm0.9 & 1.4\ud{1.2}{1.4} & -- & 0.28 & 0.52/ 6 (0.79) & PL\\ 
  27 & 1.5\ppm0.9 & (1.5) & -- & 0.08 & 0.95/ 8 (0.37) & PL\\ 
  28 & 0.4\ppm0.9 & (1.5) & -- & 0.05 & 2.26/ 7 (0.03) & PL\\ 
  29 & 3.3\ppm0.8 & 0.3\ud{0.9}{1.0} & -- & 0.75 & 0.34/ 5 (0.89) & PL\\ 
  30 & 3.0\ppm0.8 & 1.8\ud{1.3}{0.9} & -- & 0.29 & 0.60/ 5 (0.70) & PL\\ 
  31 & 3.0\ppm0.9 & 1.4\ud{0.9}{0.8} & -- & 0.30 & 0.36/ 7 (0.93) & PL\\ 
  32 & 1.5\ppm0.9 & (1.5) & -- & 0.20 & 0.39/ 7 (0.91) & PL\\ 
  33 & 2.2\ppm0.9 & (1.5) & -- & 0.21 & 0.47/ 7 (0.86) & PL\\ 
  34 & 2.7\ppm0.8 & 1.6\ud{1.2}{1.1} & -- & 0.31 & 0.56/ 5 (0.73) & PL\\ 
  35 & 4.1\ppm0.8 & 3.8\ud{0.7}{0.7} & -- & 0.51 & 0.63/ 8 (0.85) & PL\\ 
  36 & 3.8\ppm0.9 & 2.8\ud{0.7}{0.7} & -- & 0.45 & 1.66/ 7 (0.11) & PL\\ 
  37 & 2.0\ppm0.8 & (1.5) & -- & 0.09 & 2.25/ 6 (0.04) & PL\\ 
  38 & 3.3\ppm0.9 & 3.9\ud{2.1}{2.0} & -- & 0.33 & 1.07/ 6 (0.38) & PL\\ 
  39 & 3.0\ppm0.9 & 4.3\ud{3.4}{2.2} & -- & 0.33 & 0.63/ 6 (0.70) & PL\\ 
  40 & 1.7\ppm0.8 & (1.5) & -- & 0.20 & 0.66/ 5 (0.65) & PL\\ 
  41 & 0.7\ppm0.9 & (1.5) & -- & 0.01 & 2.01/ 7 (0.05) & PL\\ 
  42 & 0.4\ppm0.7 & (1.5) & -- & 0.24 & 1.37/ 4 (0.24) & PL\\ 
  43 & 1.8\ppm0.8 & (1.5) & -- & 0.49 & 0.88/ 5 (0.50) & PL\\ 
  \enddata 

  \tablecomments{The columns of this table are, from left to right:
    the source ID according to Table~\ref{tab:sources6553}, the
    background subtracted count rates in the 0.3--10\keV\ range, the
    best-fit photon index when a {\tt power law} was used (values in
    parentheses were held fixed), the best-fit effective temperature
    when a thermal model was used, and the unabsorbed \xray\ flux in
    unit of $\ee{-13}\cgsflux$ in the band 0.5--10\keV.  Note that the
    flux is not corrected for the finite aperture in this table.  This
    energy range is different from that used for the spectrum
    extraction (0.3--10\keV), for comparison purposes with previously
    published work.  The last column is the resulting \chisq-statistic
    and the model used (PL: power law, RS: Raymond-Smith plasma).  All
    fits have been performed with a fixed value of the galactic
    absorption, \nhtt=0.35.}

  \tablenotetext{a}{The MOS1/2 count rates are respectively
    8.5\ppm0.8 and 6.6\ppm0.7 counts per ksec in the
    0.3-10\keV\ range.}
\end{deluxetable*}

In the reduced image of NGC~6553, the {\tt wavdetect} algorithm
detected 43 \xray\ sources on the pn camera within the field of the
observation.  Table~\ref{tab:sources6553} lists all the sources
detected with their positions, their statistical uncertainty and their
detection significance, while Table~\ref{tab:spec6553} provides the
results of the spectral fits and the fluxes.  The spectral analysis of
candidate qLMXBs and of the sources for which a PL model does not
describe the data is detailed below.

As can be seen in Figure~\ref{fig:6553}, the proximity (17\farcs5) of
source \#15 to source \#5 requires to adapt the extraction radius
prior to the spectral analysis.  More specifically, instead of
excluding the default 40\arcsec\ around nearby sources, the exclusion
radii of source \#15 is chosen to be 10\arcsec\ (59\% of encircled
count fraction, ECF) for the count extraction of \#5, and the
exclusion radii of source \#5 is chosen to be 12\farcs5 (68\% of ECF)
for the count extraction of source \#15.  The radii mentioned were
determined by visual inspection of the \xray\ image.  Similarly, the
{\tt wavdetect} algorithm detected two sources, \#11 and \#14, located
16\arcsec\ apart near the edge of the pn camera.  In this case, the
exclusion radii of \#11 and \#14 are set to 15\arcsec\ and 10\farcs5,
when considering the extraction region of sources \#14 and \#11,
respectively.

\subsection{Source \#1 - XMMU~J180846$-$260043}
\label{sec:nonQLMXB1}
This \xray\ source has the largest pn count rate of the observation,
$0.197\pm0.004\unit{cts\persec}$.  Nevertheless, it is not affected
pile-up (accumulation of photon events within a single CCD time frame,
73$\unit{ms}$) since its count rate per detector frame is then
$\sim0.014$.

The bright \xray\ source corresponds to the \rosat\ source
1RXS~J180847.7-260037 \citep{voges99} and to the IR star 2MASS
J18084678-2600417 (with probability of association 99.999\%).  It is
also named, HD~315209, as listed in the Tycho Reference Catalog with
the proper motion $+14.2\ppm2.4\masperyr$ in R.A. and
$-9.0\ppm1.7\masperyr$ in decl. \citep{hog98,hog00}.  The star has a
spectral type K5V and is located at a distance of 33\pc, deduced from
spectroscopic observations \citep{riaz06}.

The \xmm\ source is adjacent to a bad column on the pn camera.
Therefore, the resulting astrometry is likely to be offset in the
direction opposite to the bad column, i.e., toward increasing right
ascension.  Therefore, we provide the position obtained with the MOS2
camera -- R.A.=18$^{\rm h}$08$^{\rm m}$46.7$^{\rm s}$ and ${\rm
  decl.}=-26\deg 00\arcmin 42\farcs7$.

The high-S/N spectrum of this star is not well described by an
absorbed PL: \Chisq{1.80}{120}{1.7\tee{-7}}.  Removing the absorption,
i.e., setting \nh=0, also leads to a non-acceptable fit with ${\rm
  n.h.p.}\sim\ee{-60}$. Similarly, NS H-atmosphere, thermal
bremsstrahlung and blackbody models do not fit properly the data,
leading to ${\rm n.h.p.} <\ee{-10}$, whether or not galactic
absorption is included with the models fitted.  A Raymond-Smith (RS)
plasma model provides improved fit statistics with ${\rm n.h.p.}
\approx\ee{-3}$ but this is still not a statistically acceptable fit.
A two-temperature RS model, often used to represent stellar coronal
spectra \citep{dempsey93}, satisfactorily describes the spectrum,
\Chisq{1.01}{11.5}{0.44}, with best-fit temperatures $\kteff_{,1} =
0.35\ud{0.05}{0.12}\keV$ and $\kteff_{,2} = 2.80\ud{5.75}{0.67}\keV$.
With this model, the best-fit hydrogen column density is $\nhtt =
0.09\pm0.04$, consistent with a small distance to this star which lies
in the foreground of NGC~6553.

\subsection{Source \#5 - XMMU~J180935$-$255554}
\label{sec:nonQLMXB5}
This source, located within the GC tidal radius, is likely associated
with the star HD~165973 of spectral type G8III \citep{houk88}.  The
2MASS name of this star is 2MASS J18093564-2555537 (99.999\% chance of
association with the \xray\ source \#5).  It has a proper motion of
$+12.9\ppm1.9\masperyr$ in R.A. and $-3.5\ppm1.6\masperyr$ in
decl. \citep{hog98,hog00}.  No parallax measurement is available for
this star.

The \xray\ spectrum cannot be fit with the NS H-atmosphere, PL,
blackbody and thermal bremsstrahlung models.  All of these models lead
to ${\rm n.h.p.}<0.006$.  However, an RS plasma model provides a
statistically acceptable fit with \Chisq{0.75}{13}{0.71}.  The
best-fit temperature is $\kteff=0.37\ud{0.07}{0.12}\keV$ and the
best-fit hydrogen column density is $\nhtt<0.26$ (90\% confidence
upper limit).

\subsection{Source \#9 - \sourcenineB}
\label{sec:source9}

The source \#9 has a soft thermal spectrum that is found to be
acceptably fit by the {\tt nsatmos} model: $\kteff = 80\ud{21}{12}\eV$
and $\rns = 15.4\ud{6.3}{6.0}\km$ for $\mns=1.4\msun$, with
\Chisq{1.50}{32}{0.03}.  The spectrum is also acceptably fit with a RS
plasma model: $\kteff = 0.87\ud{0.15}{0.09}\keV$,
$\nhtt=0.19\ud{0.07}{0.08}$ and $Z=0.10\ud{0.16}{0.06}$ with
\Chisq{1.20}{30}{0.21}.  A Kolmogorov-Smirnov test (probability
$p=0.22$) shows no evidence of variability.

The infrared (IR) source, \TwoMassnineB, is located at a distance of
1\farcs51 from source \#9 (after correction) and is identified as the
probable counterpart of the candidate qLMXB, with a probability of
association of 99.86\%.  It has magnitudes in the $J$, $H$ and $K$ IR
bands of $m_{J}=10.146\left(15\right)$, $m_{H}=9.796\left(23\right)$,
and $m_{K}=9.723\left(31\right)$, where the numbers in parentheses
indicate the uncertainties on the magnitudes.  The \xray\ to $J$-band
flux ratio of the association is $\fxfj \approx 2\tee{-3}$ and is
consistent with that of a coronally active star.  The possibility that
source \#9 is a qLMXB with a giant star in NGC~6553 cannot be
discarded on this basis.

However, in the color-magnitude diagram of NGC~6553
\citep{ortolani95}, the $V$-band magnitude of this star in the NOMAD
Catalogue \citep{zacharias05} lies $\sim5$ mag above the tip of the
red giant branch. This observation strongly disfavors the giant star
hypothesis, since the star is $\sim100\times$ brighter than the
brightest giants in the cluster.  In addition, the location of the
system outside the tidal radius of the cluster also disfavors the
qLMXB classification.  Therefore, we conclude that source \#9 is not a
candidate qLMXB at the distance of the GC.

\subsection{Candidate qLMXB, source \#3 - \sourcethree}
\label{sec:qlmxb3}
\subsubsection{\xmmlong\ observation}

\begin{figure}[t]
  \centerline{~\psfig{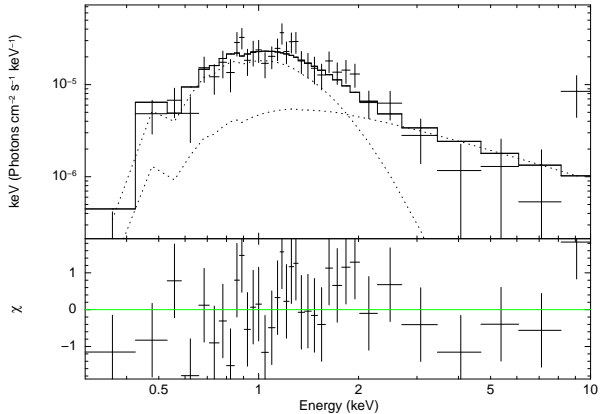}~}
  \bigskip
  \caption[]{\label{fig:6553-X3} \xmm-pn spectrum of the candidate
    qLMXB \sourcethree\ in NGC~6553.  The two components (H-atmosphere
    {\tt nsatmos} and PL {\tt powerlaw}) are shown in dashed
    lines.  They are both affected by interstellar absorption modeled
    with {\tt wabs} and the hydrogen column density \nhtt=0.35.  The
    PL dominates the spectrum at $E\approxgt2\keV$.  }
\end{figure}

The \xray\ spectrum of this source is acceptably fit
(\Chisq{1.44}{33}{0.05}) with {\tt nsatmos}, for which the parameters
are: $\rns = 6.7\ud{0.9}{1.2}\km$ and $\kteff = 137\ud{20}{18}\eV$,
for $\mns=1.4\msun$.  However, an additional PL component is required
to account for the high-energy tail of the spectrum, as evidenced by a
low f-test probability (prob.=0.002).  Adding the PL component to {\tt
  nsatmos} results in the following improved fit :
\Chisq{0.97}{31}{0.52}, with $\rns = 6.3\ud{2.9}{1.0}\km$
($\mns=1.4\msun$), $\kteff = 135\ud{25}{42}\eV$ and
$\Gamma=2.1\ud{1.5}{0.9}$ (Figure~\ref{fig:6553-X3}).  The best-fit
projected radius $\rinfty$ and temperature are consistent with the
qLMXB classification.  There is no evidence of variability on the
timescale of the observation as evidenced by a Kolmogorov-Smirnov test
(probability $p=0.25$).

To improve the statistics, the MOS1 and MOS2 extracted spectra are
added in \emph{XSPEC} and a simultaneous spectral fit is performed.  A
multiplicative constant is added to the {\tt nsatmos + powerlaw} model
to verify that there is no discrepancy between the pn spectrum and the
MOS1/2 spectra normalizations.  The multiplicative constant is fixed
to $c=1$ for pn and fit for the MOS1/2 data.  It is found that $c_{\rm
  MOS1}=1.08\ud{0.20}{0.18}$ and $c_{\rm MOS2}=0.86\ud{0.17}{0.16}$,
both consistent with unity.  Consequently, no discrepancy is observed
and the simultaneous fit is performed with the values fixed at $c=1$
for all three spectra.  The best-fit {\tt nsatmos + powerlaw}
parameters are listed in Table~\ref{tab:rescore}.  The NS parameters
and the unabsorbed luminosity (thermal component $L_{\rm
  X,th}=4.2\tee{32}\cgslum$, and PL component $L_{\rm
  X,PL}=2.1\tee{32}\cgslum$, in the 0.5--10\keV\ range) are within the
expected range of values for qLMXBs.

The PL component accounts for 33\% of the total unabsorbed flux.  Such
a strong PL component, if not intrinsic to the source, could suggest
that, as observed for the candidate qLMXB in the core of NGC~6304
\citep{guillot09b}, multiple \xray\ sources are unresolved in the
core.  Running the detection algorithm {\tt wavdetect} on the MOS1/2
images (because of the smaller pixel size than the pn camera) does not
reveal the presence of multiple resolved sources in the
core. Nonetheless, visual inspection of the EPIC images seems to
suggest that the spatial count distribution deviates from that of a
point source, favoring the hypothesis that nearby sources are
unresolved by \xmmlong.

\begin{figure}[t]
  \centerline{~\psfig{file=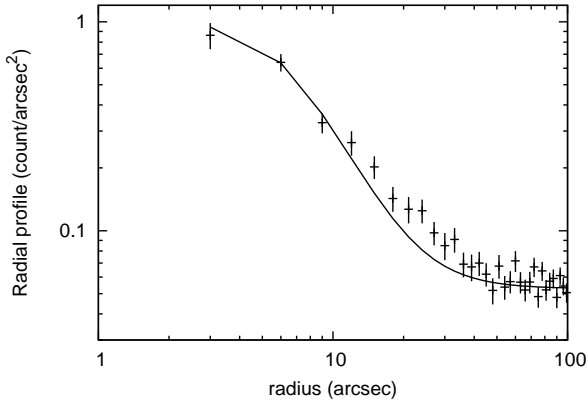,width=8cm,angle=0}~}
  \bigskip
  \caption[]{\label{fig:6553-X3-PSF} \xmm\ surface brightness as a
    function of radius of the candidate qLMXB in the core of NGC~6553.
    The line indicates the best-fit pn-camera point-spread function at
    the position of the source.  Counts in excess of the telescope PSF
    at $\sim15\arcsec$--$30\arcsec$ are evidence of the presence of one (or
    more) unresolved sources.  This is marginally supported by the
    statistic of the PSF fit: $\chisq=51.5$ for 31 d.o.f., with
    n.h.p.=1.2\%.}
\end{figure}

A more quantitative analysis is performed using the \emph{XMMSAS} task
{\tt eradial}.  By comparing the spatial distribution of counts to the
telescope PSF (Figure~\ref{fig:6553-X3-PSF}), there is marginal
evidence that the source differs from a point source (\chisq=51.5 for
31 degrees of freedom (dof), with ${\rm n.h.p}=1.2\%$).  In other
words, there is marginal support to the hypothesis that the qLMXB
candidate is confused with nearby sources (at
$\sim15\arcsec$--$30\arcsec$) in the \xmm\ data.  For this fit and for
Figure~\ref{fig:6553-X3-PSF}, the radial extent of the PSF is cut off
at 100\arcsec, the distance of the next closest \xray\ source
detected.  However, the fit statistics are highly dependent on the
radial extent of the PSF.  For example, choosing instead a cut off at
30\arcsec\ leads to \chisq=30.1 for 9 dof, with ${\rm n.h.p}=0.04\%$,
which is strong evidence for the presence of nearby unresolved
sources.

The hypothesis that the PL component is due to nearby sources is
further tested by using a spectrum created from counts located within
12\arcsec\ of the source center, therefore decreasing the
contamination from possible nearby sources at $>12\arcsec$, as
evidenced by the PSF analysis above.  However, this method reduces the
S/N since only 62\% of the ECF is included within 12\arcsec\ at
1.5\keV.  This spectrum has a count rate of 15\ppm1\unit{cts\perksec}.
Using the {\tt nsatmos} model alone, the best-fit parameters, $\rns =
6.4\ud{1.8}{0.8}\km$ (for $\mns=1.4\msun$) and $\kteff =
135\ud{19}{27}\eV$, are consistent with the best-fit parameters
mentioned above, as is the unabsorbed flux $\Fx =
1.49\tee{-13}\cgsflux$.  The statistically acceptable fit,
\Chisq{1.73}{12}{0.05}, does not require a PL component, as
demonstrated by the high F-test probability (prob. = 0.31).  Adding a
PL component leads to consistent NS parameters and flux.  Also, the
fraction of the PL to the total flux is consistent with zero at the
$1\sigma$ level.  This provides additional support to the claim that
the PL component represents the emission of nearby contaminating
sources at angular distances $>12\arcsec$ from the qLMXB candidate in
the core of NGC~6553.

Finally, the second \xmmlong\ serendipitous source catalog
\citep[2XMMi;][]{watson09} reports two sources located 20\farcs7
apart, at the positions R.A.=18$^{\rm h}$09$^{\rm m}$16.5$^{\rm s}$
and ${\rm decl.}=-25\deg 54\arcmin 26\farcs1$ for the brightest one,
coincident with the qLMXB and, R.A.=18$^{\rm h}$09$^{\rm m}$18.0$^{\rm
  s}$ and a fainter nearby source at ${\rm decl.}=-25\deg 54\arcmin
27\farcs5$.  The discrepancy between 2XMMi and the detections in this
work may reside in the algorithm used.  While we performed the
detection with {\tt wavdetect}, the 2XMMi sources are detected with a
sliding box algorithm.  In addition, we used the pn data only, while
the 2XMMi catalog is constructed with a pn+MOS1/2 mosaic.

\begin{figure}[t]
  \centerline{~\psfig{file=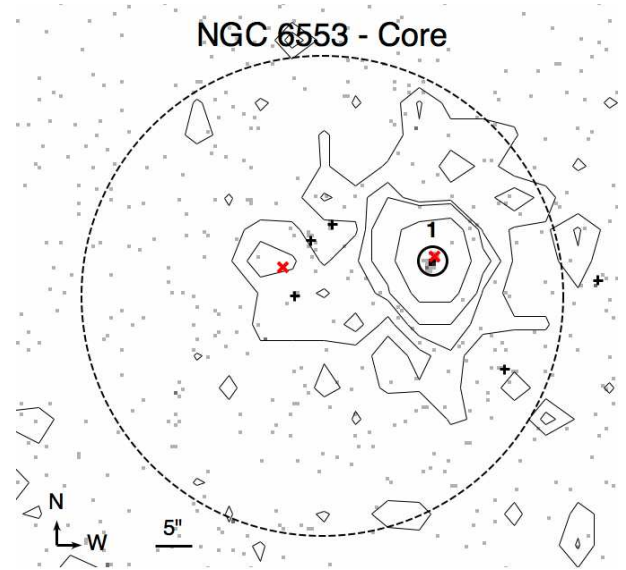,width=8cm,angle=0}~}
  \bigskip
  \caption[]{\label{fig:ngc6553core} \xmm\ contours plotted over the
    \chandra\ ACIS-S3 image.  The small circle is the candidate qLMXB
    detected with \chandra. The large dashed circle represents the
    $0\farcm55$ core radius of the GC.  The red crosses '$\times$'
    show the positions of the two 2XMMi sources.  Finally, the five
    black '$+$' signs show the possible sources that have
    low significance in this short $\sim5\ksec$
    \chandra\ observations. }
\end{figure}

Overall, the above analysis supports the hypothesis that one or more
nearby sources produce the observed PL component.  However, only the
angular resolution of the \chandralong\ offers the possibility to
ascertain that the above claim is correct, by spatially and spectrally
differentiating multiple sources.  Archived \chandra\ data of NGC~6553
are analyzed and the results are presented in the following
subsection.

\subsubsection{\Chandra\ observations of NGC~6553}

The following pertains solely to the confirmation of the candidate
qLMXB and makes use of \chandra's angular resolution to understand the
core-source emission.  A systematic detection and spectral analysis of
all \xray\ sources are not reported here.

Only one source is detected with {\tt wavdetect} in the core of the
GC, with a background subtracted count number of $46.8\ppm6.9
\unit{cts}$.  The position of CXOU~180916.48$-$255426.58 is consistent
with the corrected \xmm\ position of the candidate qLMXB.  The
difference between the \chandra\ and \xmm\ positions, 1\farcs4, is
within the accuracy of the \chandra\ systematic uncertainty (0\farcs6)
and the \xmm\ residual error after the correction (1\farcs1, see
Section~\ref{sec:astro}).  Therefore, the \chandra\ source is
consistent with being the same \xray\ source as
\sourcethree. Figure~\ref{fig:ngc6553core} shows the \chandra\ image
of the GC core, with \xmm\ contours and the two 2XMMi catalog
sources.

A second possible source, with a $2.1\sigma$ significance
($1.28\ppm0.54\unit{cts\perksec}$), lies 17\arcsec\ from the core
candidate qLMXB, and close ($\sim5\arcsec$) to the faint 2XMMi source.
An additional one, ($\sim4\arcsec$) away from the faint 2XMMi source
and 19\farcs5 distant from the qLMXB candidate, has a significance of
$1.6\sigma$.  Other faint possible sources, with low-significance
($1.6\sigma$, $1.7\sigma$ and $2.2\sigma$) are located at distances
14\farcs8, 17\farcs7 and 22\farcs7, respectively (see
Figure~\ref{fig:ngc6553core}).  The extended emission observed in
Figure~\ref{fig:6553-X3-PSF} could be explained by the presence of
other sources in the GC core.  Unfortunately, this short
\chandra\ observation does not permit a confident detection of all the
sources in the core.  Furthermore, restricting counts to the
3--10\keV\ or the 2--10\keV\ energy ranges in {\tt wavdetect} does not
lead to the detection of any nearby source, suggesting that this
excess of counts above 2\keV\ is not due to a single source, but due
to multiple fainter sources.  A longer high angular resolution
observation of the central region of NGC~6553 would certainly shed
light on the PL tail observed in the \xmm\ spectrum of the candidate
qLMXB

\begin{deluxetable*}{lrcccccr}
  \tablecaption{\label{tab:rescore} Spectral analysis of the core candidate qLMXB}
  \tablewidth{0pt}
  \tabletypesize{\scriptsize}    
  \tablecolumns{7}
  \tablehead{
    \colhead{Detector} & \colhead{Model}  & \colhead{\rns} & \colhead{\kteff} & \colhead{$\Gamma$} &
    \colhead{$F_{\rm X}$} & \colhead{PL Flux} & \colhead{\chisqrnu/d.o.f. (prob.)}\\
    \colhead{} & \colhead{} & \colhead{(km)} & \colhead{(eV)} & \colhead{} &
    \colhead{} & \colhead{(\%)} & \colhead{}}
  \startdata
  EPIC-pn (12\arcsec) & NSATMOS + PL & 6.4\ud{3.1}{1.1} & 131\ud{28}{42} & 2.0\ud{0.8}{3.6} & 1.6\ud{6.9}{1.2} & 23\ud{48}{23} & 1.37/10 (0.18) \\
  EPIC-pn+MOS1/2      & NSATMOS + PL & 6.3\ud{2.3}{0.8} & 136\ud{21}{34} & 2.1\ud{0.5}{0.8} & 1.5\ud{4.2}{0.7} & 33\ud{32}{22} & 0.90/54 (0.68) \\
  EPIC+ACIS-S3 \tablenotemark{a} & NSATMOS + PL  & 6.4\ud{2.1}{0.8} & 134\ud{21}{34} & 2.1\ud{0.5}{0.7}  & 1.5\ud{4.3}{0.7} & 32\ud{35}{22} & 1.02/58 (0.44)  \\
  ACIS-S3             & NSATMOS       & 6.6\ud{3.2}{1.3} & 134\ud{35}{39} &         --     & 1.0\ud{4.6}{0.3}  & -- & cstat=146.98    \\
  \enddata 

  \tablecomments{In all the fits, the galactic absorption, modeled
    with {\tt wabs}, is represented by a fixed value of
    $\nhtt=0.35$. $\Gamma$ represents the photon index of the
    power-law (PL) model.  $F_{\rm X}$ is the total unabsorbed flux,
    in units of $\ee{-13}\cgsflux$, in the range 0.5--10\keV, with
    $1\sigma$-errors. {\tt PL Flux (\%)} represents the PL
    contribution to the total flux ($1\sigma$-errors).}

  \tablenotetext{a}{For the ACIS-S3 spectrum, the {\tt
      powerlaw} normalization is set to zero.  The multiplication
    constant, set free for ACIS-S3, is $c_{\rm
      ACIS}=0.85\ud{0.43}{0.26}$.}
\end{deluxetable*}

The PL component measured with \xmm\ has an absorbed flux
$\Fx=4.9\tee{-14}\cgsflux$ in the 0.5--10\keV\ range with a photon
index $\Gamma=2.1$.  Using \emph{webPIMMs}\footnote{Available via
  HEASARC, \url{http://heasarc.gsfc.nasa.gov/Tools/w3pimms.html}}, the
expected \chandra/ACIS-S3 count rate is $4.7\unit{cts\perksec}$
(0.5--8\keV), which corresponds to 23.5 counts on this short
\chandra\ observation.  However, the \chandra\ background subtracted
count rate of an annulus ($r_{\rm in}=10\arcsec$ and $r_{\rm
  out}=30\arcsec$) centered around the core source is
$7\pm14\unit{cts}$.  The non-detection of an excess of counts around
the core source does not support the possibility that multiple diffuse
sources are responsible for the high-energy contamination of the
\xmm\ spectrum of the candidate qLMXB.  This could be explained if the
sources are variable. 

Alternatively, the detected PL component could be intrinsic to the
candidate qLMXB, but display some variability, as has been observed
for other LMXBs.  This possibility is investigated by extracting
counts in the 2--10\keV\ range. With the pn camera, a count rate of
$0.97\ppm0.31\unit{cts\perksec}$ is detected, which corresponds to
$0.36\unit{cts\perksec}$ expected on the ACIS-S3 detector using
\emph{webPIMMs}, assuming a photon index $\Gamma=2.1$.  We detect
$1.13\ppm0.47\unit{cts\perksec}$ on the \chandra\ data in the
2--10\keV\ range.  These count rates are consistent and therefore
there is no observed variability of the high-energy component
($>2\keV$), at the $<2\sigma$ level.

The \chandra\ spectrum of the candidate qLMXB in the core of NGC~6553,
i.e., image on-axis, is extracted from a 2\farcs5 radius circular
region to include more than 95\% of the source
energy\footnote{\Chandra\ Observatory Proposer Guide, Chap. 6, v11.0,
  Jan 2009}.  For the purpose of the count rate calculation only, the
background is extracted using a circular region of 25\arcsec\ radius
around the source , excluding 5\arcsec\ around the qLMXB itself, which
ensures that $>99\%$ of the ECF is excluded.

\begin{figure}[t]
  \centerline{~\psfig{file=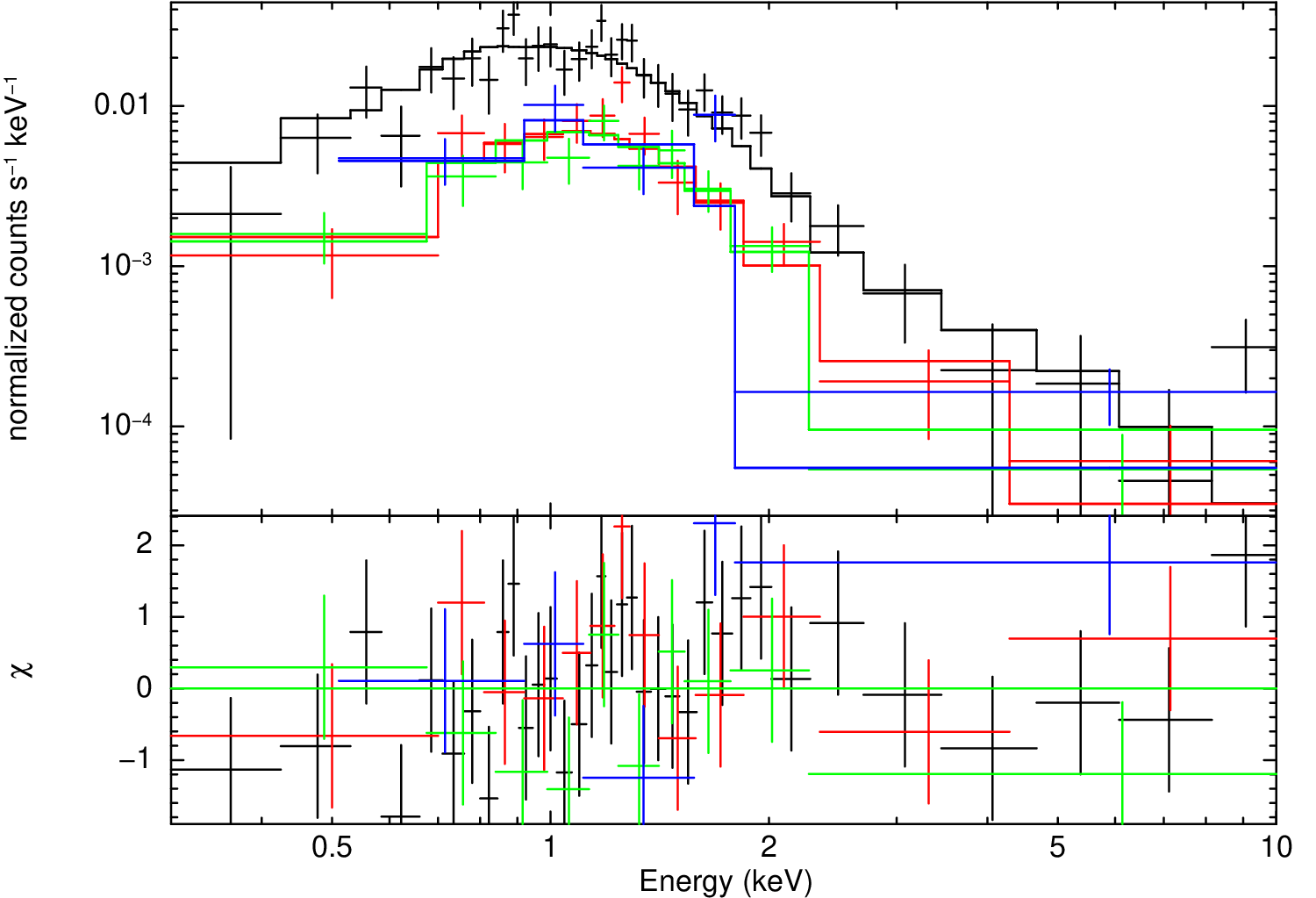,width=8cm,angle=-90}~}
  \bigskip
  \caption[]{\label{fig:6553-X3-all} \xmm-pn (black), MOS1 (red), MOS2
    (green) and ACIS-S3 (blue) spectra of the candidate qLMXB of
    NGC~6553, \sourcethree, with the best-fit {\tt nsatmos+powerlaw}
    model with {\tt wabs} galactic absorption \nhtt=0.35.  The ACIS-S3
    counts are grouped with a minimum of 10 cts per bin, due to the
    small number of counts.  The fit statistic \Chisq{1.02}{58}{0.44}
    and the multiplicative constant applied to the ACIS-S3 spectra
    $c=0.85\ud{0.43}{0.26}$ (consistent with unity) are evidence that
    the \chandra\ spectrum is consistent with the \xmm-EPIC spectra,
    within the uncertainty.}
\end{figure}

Due to the small numbers of counts, a binning of 10 counts per bin is
applied for the candidate qLMXB, which only guarantees marginally
Gaussian uncertainty in each bin.  The binned spectrum is used
simultaneously with the three EPIC spectra to evaluate the consistency
of the spectra.  All parameters are tied between the four spectra.
Only the {\tt powerlaw} norm is set to zero for the
\chandra\ observation.  Also, a multiplicative constant is added to
the model to quantify the variation between EPIC and ACIS-S3 spectra.
We find that the best-fit parameters of the four spectra are
consistent with that of the EPIC spectra alone (see
Table~\ref{tab:rescore} and Figure~\ref{fig:6553-X3-all}).  In
addition, the multiplication constant, which was left free for the
ACIS-S3 spectrum, is statistically consistent with unity, $c_{\rm
  ACIS}=0.85\ud{0.43}{0.26}$.  This is evidence that, within the
statistics of the observation, the \chandra\ spectrum is consistent
with the three EPIC spectra.

Finally, we also perform a fit of the \chandra\ unbinned spectrum
alone, using Cash-statistic \citep{cash79}.  The background,
representing $2.5\%$ of the extracted counts (1.2 counts out of 48 in
the source extraction region) can be neglected, as required by
Cash-statistic.  For comparison purposes with the \xmm\ spectral
fitting, the chosen model here is {\tt nsatmos} alone (with fixed {\tt
  wabs} absorption, distance and mass as before), which assumes that
the PL component observed in the \xmm\ spectrum is not intrinsic to
the candidate qLMXB, i.e., that in the \xmm\ spectrum, the PL tail is
due to contaminating nearby sources.  The best-fit value are $\rns =
6.6\ud{3.2}{1.3}\km$ ($\mns=1.4\msun$) and $\kteff =
134\ud{35}{39}\eV$, which are in remarkable agreement with the values
obtained from the \xmm\ observation.  The 0.5--10\keV\ unabsorbed
\xray\ flux is $\Fx = 1.0\tee{-13}\cgsflux$.  These results are also
listed in Table~\ref{tab:rescore}, for a convenient comparison with
the \xmm\ spectral fits.

\section{Discussion and Conclusion}
\label{sec:discuss}

Using the \xmm\ observation of NGC~6553, we have spectrally identified
one candidate qLMXB in the GC. The best-fit {\tt nsatmos} parameters
found are $\rns = 6.3\ud{2.3}{0.8}\km$ (for $\mns=1.4\msun$) and
$\kteff = 136\ud{21}{34}\eV$, and the best-fit PL photon index is
$\Gamma=2.1\ud{0.5}{0.8}$.  The low-S/N \chandra\ spectrum finds
consistent best-fit values within the statistic permitted by the
detected counts.  The modest precision of the NS radius measurement
obtained with these short exposures does not permit to place
constraints on the dense matter EoS.  Deeper \chandra\ exposures will
provide the S/N necessary to confirm the classification and to produce
useful uncertainty on the radius measurement for the determination of
the dense matter EoS.

The number of candidates identified in NGC~6553 (only 1) is in
agreement with that predicted from the relations
$N_{\rm qNS}\sim0.04\times\Gamma_{\rm GC}+0.2$ \citep{gendre03b} and
$N=0.993\times\Gamma_{\rm GC}-0.046$ \citep{heinke03c}, where the
authors used different normalizations for $\Gamma_{\rm GC}$; the
encounter rate of the GC is defined by $\Gamma_{\rm GC} \propto
\rho_{0}^{1.5} r_{\rm c}^{2}$ \citep{verbunt03a}.  In this last
equation, $\rho_{0}$ is the central luminosity density (in
$\unit{\lsun\,pc^{-3}}$) and $r_{\rm c}$ is the core radius (in pc,
not angular distance).  Quantitatively, using the same respective
normalizations, the number of expected qLMXBs in NGC~6553, given
$\rho_{0, {\rm NGC~6553}} = 3.84\unit{\lsun\,pc^{-3}}$
\citep{harris96}, are 0.43 and 0.34 using the two relations cited
above, respectively.  Assuming Poisson statistics, the probability of
finding one qLMXB when 0.43 (0.34) are expected is 28\%
(24\%). Therefore, our findings are not in disagreement with these two
empirical predictions.

The total unabsorbed \xmm\ flux of the candidate qLMXB,
$\Fx=1.5\ud{4.3}{0.7}\tee{-13}\cgsflux$, includes a PL component with
a 33\ud{35}{22}\% contribution to the flux.  If intrinsic to the
source, this PL component has one of the strongest contribution to the
total flux ever observed for a qLMXB in a GC \citep[others are in
  NGC~6440 and 47Tuc,][respectively]{heinke03c,heinke05}. However, we
have shown evidences for the presence of nearby contaminating sources.
This is tentatively supported by a short archived
\chandra\ observation, but the low-significance detection of the
nearby sources does not confirm with certainty the existence of
contamination.  Only a longer \chandra\ observation would confirm the
spectral classification of the bright core source and confirm the
detection of the faint neighboring core sources.

The candidate qLMXB did not display \xray\ variability on the time
scale of the observation.  Specifically, there was no evidence that
the \xmm\ integrated light curve is significantly different from a
linear distribution of counts.  Regarding the long time scale
variability, there is no significant change of flux between the
\xmm\ and \chandra\ observations.  We also searched for archived
\rosat\ observation, but no source was detected at the position of
\sourcethree\ on the PSPC survey, and no HRI observation at the center
of the GC was performed.

To conclude, the candidate in the core of NGC~6553, with $L_{\rm
  X,th}=4.2\tee{32}\cgslum$ ($d=6\kpc$), adds to the small list of
known GC qLMXB.  However, only a high enough S/N observation with
\chandra\ will allow to confirm with certitude the spectral
classification of the source, and will be able to provide constraints
on the dense matter EoS.

\acknowledgements The authors thank the anonymous referee for the comments that contributed to the improvement of this article. S.G. acknowledges the support of NSERC via the Vanier CGS program.
R.E.R. is supported by an NSERC Discovery grant.  The work by GGP was
partially supported by NASA (grant NNX09AC84G) and by the Ministry of
Education and Science of the Russian Federation (contract
11.G34.31.0001).  The authors also acknowledge the use of
\xmmlong\ observations and the use of archived data from the High
Energy Astrophysics Archive Research Center Online Service, provided
by the NASA GSFC.

\bibliographystyle{apj_8}
\bibliography{biblio}





\clearpage





\end{document}